\documentclass[preprint,12pt]{elsarticle}
\usepackage{graphicx}
\usepackage{amsmath}
\usepackage{amssymb}
\usepackage{upgreek}

\newcommand{\be}{\begin{equation}}
\newcommand{\ee}{\end{equation}}
\newcommand{\hattheta}{\hat{\boldsymbol{\theta}}}

\newcommand{\bdsigma}{\boldsymbol{\sigma}}

\begin{document}

\begin{frontmatter}

\title{Electromagnetic Momentum Conservation in media}

\author{Iver Brevik}
\author{Simen {\AA}. Ellingsen}
\address{Department of Energy and Process Engineering, Norwegian University of Science and Technology, N-7491 Trondheim, Norway}

\begin{abstract}

That static electric and magnetic fields can store momentum may be perplexing, but is necessary to ensure total conservation of momentum. 
Simple situations in which such field momentum is transferred to nearby bodies and point charges have often been considered for pedagogical purposes, normally assuming vacuum surroundings. 
If dielectric media are involved, however, the analysis becomes more delicate, not least since one encounters the electromagnetic energy-momentum problem in matter, the `Abraham-Minkowski enigma', of what the momentum is of a photon in matter.
We analyze the momentum balance in three nontrivial examples obeying azimuthal symmetry, showing how the momentum conservation is satisfied as the magnetic field decays and momentum is transferred to bodies present. In the last of the examples, that of point charge outside a dielectric sphere in an infinite magnetic field, we find that not all of the field momentum is transferred to the nearby bodies; a part of the momentum appears to vanish as momentum flux towards infinity. We discuss this and other surprising observations which can be attributed to the assumption of magnetic fields of infinite extent. 
We emphasize 
how formal arguments of conserved quantities cannot determine which energy-momentum tensor is more ``correct'', and each of our conservation checks may be performed equally well in the Minkowski or Abraham framework.

\end{abstract}

\end{frontmatter}

\section{Introduction}

It may appear surprising at first sight that there is an
electromagnetic field momentum with density
\begin{equation}
{\bf g=D\times B} \label{1}
\end{equation}
for crossed electric and magnetic fields in a vacuum field even under
static conditions. This somewhat striking property of classical
electromagnetic theory has actually served as  a popular
demonstration example to show how the law of momentum conservation
works for a closed system.

\begin{figure}[ht]
  \begin{center}
    \includegraphics[width=3in]{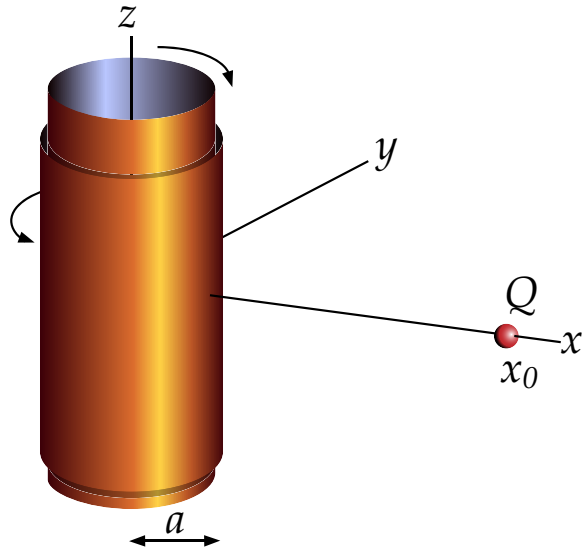}
    \caption{Two metallic cylinders rotating in opposite senses. }
    \label{fig1}
  \end{center}
\end{figure}

For illustration, let
consider the setup of
Fig.~\ref{fig1}, where there are two infinitely long metallic concentric
cylinders carrying opposite charge with nearly the same radius $a$, one cylinder fitting
inside the other.
The following argument is adopted from the exposition in Aharonov and Rohrlich's book \cite{aharonov05}.
Initially the cylinders are made rotate in opposite senses, with constant and opposite
angular velocities. At a distance $x=x_0$ from the common axis of the
cylinders there is placed a heavy particle with charge $Q$. Assume
that the cylinders gradually stop rotating, because of friction
between them. What happens to the momentum balance?

Initially, if $x_0 \gg a$ the electric field component $E_x$
within the cylinder volume is practically the same as on the $z$
axis (SI units assumed)
\begin{equation}
E_x=-\frac{Q}{4\pi \varepsilon_0}\frac{x_0}{(x_0^2+z^2)^{3/2}}.
\label{2}
\end{equation}
(We assume here that the metallic walls have no influence on the
electric field on the inside. This is clearly an unphysical
assumption, but it has no importance for the argument. As an
alternative, one might assume that the magnetic field were
produced by external sources not influencing the electric field at
all.) While the cylinders rotate
at constant velocity the
magnetic field inside is constant, $B_z=B_0$. The $y$ component of
momentum is thus $g_y=-\varepsilon_0 E_xB_0$, which implies that that
the initial total field momentum $G_y$ is
\begin{equation}
G_y=\int_{\rm cylinder} g_y
dV=\frac{1}{4}QB_0x_0a^2\int_{-\infty}^\infty
\frac{dz}{(x_0^2+z^2)^{3/2}}=\frac{QB_0a^2}{2x_0}. \label{3}
\end{equation}

Assume now that at $t=0$ the cylinder movement begins to slow down, for example due to friction. The magnetic field decays and eventually vanishes when the cylinders come to a halt. During this decay period
an azimuthal induced electric field component
$E_\theta^{\rm induced}=-\dot{B}a^2/(2r)$ is induced in the exterior region. The total
impulse imparted to $Q$ by the force $F_y=QE_y^{\rm induced}$ at $r=x_0$ is thus
\begin{equation}
P_y=\int_0^\infty F_y dt=\frac{QB_0a^2}{2x_0}. \label{4}
\end{equation}
The expressions (\ref{3}) and (\ref{4}) are the same, thus
confirming  conservation of the  $y$ component of total
momentum.

Variations of the
above
argument, as already mentioned, are often
presented in connection with the foundations of electrodynamics -
cf., for instance, Ref.~\cite{feynman64}. These considerations
actually have a close bearing also on the Aharonov-Bohm effect and
the concept of ``hidden momentum" in electrodynamics (cf., for
instance,  Refs.~\cite{shockley67}; a recent feature article is
Ref.~\cite{batelaan09}).

The question now emerges -- and this is the topic to be considered
in the present paper -- in which ways does the argument above
change if the test region is filled with a dielectric medium? As
we will see, analysis of such situations can be quite
subtle. That some complications necessarily have to occur, is
evident already from our initial expression (\ref{1}) for the
momentum density: in the presence of a medium that expression is
no longer trivial but becomes the same as the Minkowski
expression \cite{mikowski08}, just one of the alternatives that is frequently
discussed in the current literature. Other alternatives, such as
the Abraham expression \cite{abraham09} which is also well known, would correspond
to different expressions for $\bf g$ and thus modify the momentum
balance calculation. The modification that we are considering, is
thus closely connected with the problem what has turned out to be
known as  the `Abraham-Minkowski problem'.

A nice introduction
to the Abraham-Minkowski problem
is found in M{\o}ller's book \cite{moller72}. A review article was given by one of the present authors some years ago \cite{brevik79}. Another review, up to 2007, is given by Pfeifer {\it et al.} \cite{pfeifer07}. Some more recent papers are Refs.~\cite{brevik09,mansuripur09,hinds09,barnett10,bradshaw10,barnett10a,mansuripur10,brevik10}.

We will consider three different examples of momentum conservation checks in the presence of dielectric media, in order of increasing complexity. First, we consider a natural extension of the introductory example given above, that of a point charge in the vicinity of a magnetic field which is restricted to a cylindrical volume, all embedded in a dielectric medium of spatially constant permittivity. 

In the light of this first example we discuss an important point, namely that, contrary to the claims by some authors in the past, such formal considerations of conserved quantities cannot be used to determine the ``correctness'' of one formulation of electromagnetic momentum over another, for example the famous formulations due to Abraham and Minkowski mentioned above. 
Next, in Sect. 3 we consider the case of an infinite dielectric cylinder with a parallel line charge placed on the outside. The final, and most involved, example, is the geometry of point charge in the vicinity of a dielectric sphere  placed in a magnetic field of infinite extent. In this final example we observe that not all of the field momentum is in fact transferred to the sphere and point charge as the magnetic field decays: a part of it is transmitted, at least formally, as momentum flux towards infinity. The point is a subtle one, however, and paradoxes arise due to the unphysical assumption that the magnetic field fills all of space, which we discuss before concluding.

\section{Example I: Point charge near confined magnetic field in a medium}\label{sec:Ia}

\begin{figure}[ht]
  \begin{center}
    \includegraphics[width=3in]{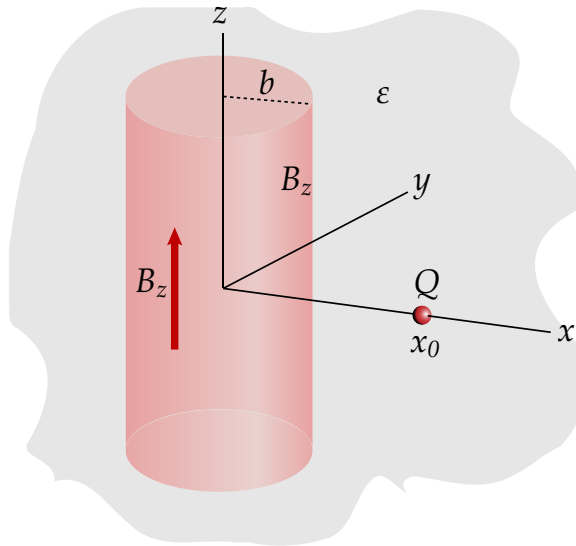}
    \caption{Charge near magnetic field confined to cylindrical volume, all embedded in an infinite medium of permittivity $\varepsilon$. }
    \label{fig2}
  \end{center}
\end{figure}

Consider the situation sketched in Fig.~\ref{fig2}: a point particle of charge $Q$ is placed in the exterior vicinity of a long cylindrical volume of radius $b$ wherein a uniform and axially directed magnetic field $\mathbf{B}=B_0 \hat{\mathbf{z}}$ is imposed. At a point in time the magnetic field starts decaying, and we study how the momentum of the electric and magnetic fields in the cylindrical volume is transferred to the particle. This is essentially the same system as in the introductory example, but in the presence of an infinite medium. We assume there is zero magnetic field at radii greater than $b$. For generality let the system be embedded in an infinite medium of permittivity $\varepsilon$ (we define the permittivity $\varepsilon$ as nondimensional, so that the static constitutive relation reads ${\bf D}=\varepsilon \varepsilon_0 \bf E$).

To facilitate calculations, let us place the system in a large perfectly conducting cylindrical ``box'' of length $L$ and radius $R$. We will assume eventually that $L$ and $R$ are much greater than other length scales in the geometry.
The electric potential at the conducting surfaces at $r=R$ and $z=\pm L/2$ is $\Phi=0$. The charge $Q$ sits on the $x$ axis at position $r'=r_0=x_0,\, \theta'=0, \,z'=0$. We initially assume the particle to lie outside the area of the magnetic field, $r_0>b$. The introduction of the large external cylindrical ``box'' enables us to deal with a completely closed physical system.

One may start the calculation by use of the Green function $G$ for the Poisson equation, which satisfies
\begin{equation}
\nabla^2 G({\bf r, r'})=-\frac{4\pi}{r}\delta (r-r')\delta (\theta-\theta')\delta (z-z'). \label{5}
\end{equation}
If $G$ is known, the potential follows from
\begin{equation}
\Phi=\frac{1}{4\pi \varepsilon \varepsilon_0}G. \label{6}
\end{equation}

The solution of the potential problem in the case of a vacuum filled closed cylinder of length $L$ and radius $R$ can be found, for instance, in Ref.~\cite{jackson99}. When generalized to a medium, the solution can be written
\begin{align}
 \Phi(r,\theta,z; r_0) =&\frac{Q}{\pi \varepsilon \varepsilon_0L}\sum_{m=-\infty}^\infty \sum_{p=1}^\infty
 \cos(m\theta)
 \sin(pqz)\sin \frac{p\pi}{2} \notag  \\
& \times I_m(pqr_<)\left[K_m(pqr_>)-I_m(pqr_>)\frac{K_m(pqa)}{I_m(pqa)} \right]. \label{7}
\end{align}
Here $I_m$ and $K_m$ are modified Bessel functions,
$r_{>}$ ($r_{<}$) is the greater (smaller) of $r$ and $r_0$,
and we have defined $q$ as
\begin{equation}
q=\frac{\pi}{L}. \label{8}
\end{equation}
We require the $r$ and $\theta$ components of the electric field ${\bf E=-\nabla} \Phi$ for radii $\leq b$ (i.e. $r<r_0$):
\begin{align}
  E_r(r<r_0)=&-\frac{Q}{\varepsilon\varepsilon_0 L^2}\sum_{m=-\infty}^\infty \sum_{p=1}^\infty p
  \cos(m\theta)
  \sin(pqz)\sin \frac{p\pi}{2} \notag \\
&  \times I_m'(pqr)\left[K_m(pqr_0)-I_m(pqr_0)\frac{K_m(pqR)}{I_m(pqR)}\right], \label{9} \\
 E_\theta(r)=&\frac{Q}{\pi \varepsilon
\varepsilon_0L}\frac{1}{r}\sum_{m=-\infty}^\infty \sum_{p=1}^\infty
m\sin(m\theta)\sin(pqz)\sin\frac{p\pi}{2} \notag \\
&\times I_m(pqr)\left[
K_m(pqr_0)-I_m(pqr_0)\frac{K_m(pqR)}{I_m(pqR)}\right]. \label{11}
\end{align}
A prime means differentiation with respect to the whole argument.

We
shall calculate the  $y$ component of the total Minkowski momentum,
similarly as above (from the symmetry it is obvious that the electromagnetic
field can have no net momentum in the $x$ or $z$ directions). This momentum is
\begin{equation}
G_y^M=B_0\int_0^b rdr \int_0^L dz \int_0^{2\pi}d\theta(-D_r\cos
\theta+D_\theta\sin \theta). \label{12}
\end{equation}
The integrals may be processed by standard relationships including
\begin{align}
  \int_0^{2\pi}d\theta{}\cos\theta \cos m\theta
  =&\int_0^{2\pi}d\theta{}m\sin\theta \sin m\theta
  =\pi(\delta_{m1}+\delta_{m,-1}), \label{13}\\
 \int_0^L\sin(pqz)dz=&\frac{2}{pq}\sin^2\frac{p\pi}{2} \label{14}
\end{align}
and
\begin{align}
  \int_0^b dr I_1(p q r) =& \frac1{pq}[I_0(pqb)-1];\notag \\
  \int_0^b dr rI'_1(p q r)=&\frac1{p^2q^2}[1-I_0(p q b)+pqbI_1(pqb)],
\end{align}
noting that $I_1(x)=I_{-1}(x)$ and $K_1(x)=K_{-1}(x)$.
After some  algebra we obtain the expression, inserting $r_0=x_0$,
\begin{align}
 G_y^M=&\frac{4QB_0 b}{\pi}\sum_{p=1}^\infty \frac{1}{p}\sin^3\frac{p\pi}{2}I_1(pqb)K_1(pqx_0) 
 \left[1-\frac{I_1(pqx_0)}{I_1(pqR)}\frac{K_1(pqR)}{K_1(pqx_0)}\right]. \label{15}
\end{align}
This expression gets contributions from $p=1,3,5...$. It is therefore convenient to introduce a new integer $l=(p+1)/2$ such that all values $l=1,2,3,...$ contribute. In terms of $l$ and $q=\pi/L$ we then get 
\begin{align}
 G_y^M=&\frac{4QB_0 b}{\pi}\sum_{l=1}^\infty \frac{(-1)^{l-1}}{2l-1}\,I_1[(2l-1)qb]K_1[(2l-1)qx_0] \notag \\
& \times \left[ 1-\frac{I_1[(2l-1)q x_0]}{I_1[(2l-1)qR]}\,\frac{K_1[(2l-1)qR]}{K_1[(2l-1)q x_0]} \right]. \label{16}
\end{align}
This expression holds for all values of $b$ up to $b=x_0$.

We now take the limits $R,L \gg x_0, b$. Whichever way this combined limit is taken, one finds that the second term of the square brackets in \eqref{16} vanishes. Moreover, by using the expressions $I_1(x) \sim x/2, \, K_1(x) \sim 1/x$ which hold for small arguments $x$, we get $I_1[(2l-1)qb]K_1[(2l-1)qx_0] \sim b/(2x_0)$ for  moderate values of $l$, which is in order since the resulting $l$ series is convergent. Observing that $1-\frac1{3}+\frac1{5}-\frac1{7}+...=\pi/4$ we  obtain in this limit the very simple expression
\begin{equation}
G_y^M=\frac{QB_0 b^2}{2x_0}. \label{18}
\end{equation}
The result is independent of the large external cylindrical box as it should. Moreover it is identical to Eq.~\eqref{4}, although we have not made any assumption here that $x_0$ be much greater than $b$.

Consider now the time interval $t>0$ in which the magnetic field decays. Since the outer cylinder is far away we can assume that the induced electric field component $E^{\rm induced}_\theta$ at radius $r_0$ is not influenced significantly by the condition that it must vanish on the boundary at $r=R\gg r_0$. The induced azimuthal electric field $E_\theta^{\rm induced}$ at a radius $b < r \ll R$ is then simply found from Faraday's law as
\begin{equation}
E_\theta^{\rm induced}=-\frac{b^2}{2r}\dot{B}. \label{19}
\end{equation}
The $y$ component of the impulse imparted to $Q$ because of $E_\theta^{\rm induced}$ is, as in Eq.~(\ref{4}), the time integral of $F_y=QE_\theta$ evaluated at $r=x_0,\, \theta=0$,
\begin{equation}
P_y=\int_0^\infty F_ydt=\frac{QB_0 b^2}{2x_0}. \label{20}
\end{equation}
Equations (\ref{18}) and (\ref{20}) are in agreement, thus demonstrating the momentum balance.

We have restricted the width of the uniform initial magnetic field so that $b<x_0$. We will consider the case $b> x_0$ below in section \ref{sec:Ib}, but let us first turn to an important point regarding different proposed forms of the electromagnetic momentum in media, which we are now able to illustrate in the light of the above example.

\subsection{Abraham vs. Minkowski momentum}\label{sec:abrahamMinkowski}

We have so far
considered the Minkowski momentum only. As mentioned in the
Introduction there exist other proposals for the momentum also.
We will argue here how, in contradistinction to some claims in the literature,
one cannot use momentum balance considerations such as those we consider in this paper to determine which energy-momentum tensor is ``correct''.

The most well-known among the alternatives to Minkowski's tensor is that introduced by Abraham \cite{abraham09}.
In the Abraham tensor, the momentum density ${\bf
g}^A$ is given as
\begin{equation}
{\bf g}^A=\frac{1}{c^2}\bf (E\times H). \label{21}
\end{equation}
Equations (\ref{16}) or (\ref{18}) would in this case be replaced
by
\begin{equation}
G_y^A=\frac{1}{\varepsilon}\,G_y^M. \label{22}
\end{equation}
The simple relationship between expressions (\ref{18}) and
(\ref{20}) is thus violated, $G_y^A \neq P_y$, since $P_y=G_y^M$. The question
naturally emerges: does this imply that the Abraham form runs into
conflict with the conservation of total momentum? The answer is
no. The reason is that the Abraham energy-momentum tensor
describes a non-closed system and is  not divergence-free. There
exists an Abraham force density ${\bf f}^A$, acting on the medium
even in spatially uniform regions. Its magnitude is
\begin{equation}
{\bf f}^A=\frac{\varepsilon-1}{c^2}\frac{\partial}{\partial t}(\bf
E\times H) \label{23}
\end{equation}
It is usually called the Abraham term. Under static conditions
corresponding to $t<0$ the term plays no role at all, but for
$t>0$ when $\dot{\mathbf{B}} \neq 0$ the force (\ref{23}) has to impart a
mechanical impulse to the medium. 
The impulse thus imparted is (we assume non-magnetic medium throughout for simplicity)
\begin{align}
  \Delta \mathbf{G}^A =& \int_0^\infty dt \int_V d^3 r {\bf f}^A = -\frac{\varepsilon-1}{\varepsilon}  \int_V d^3 r (\mathbf{D}_0\times\mathbf{B}_0) \notag\\
  =& \mathbf{G}^A - \mathbf{G}^M\label{abrahamMinkowski}
\end{align}
where $V$ is the volume in which both the electric and magnetic fields are nonzero, and subscript $0$ means the value at $t=0$. 

As Eq.~\eqref{abrahamMinkowski} shows, the difference between the Abraham and Minkowski values of the initial field momentum is imparted to the integration volume during the decay period, and thus the momentum balance is ensured also in this formalism: the total momentum transferred from the initial fields to \emph{all} bodies surrounding bodies is $P_y + \Delta G^A$ which equals the initial momentum in both formalisms. Thus the momentum conservation is demonstrated also in the Abraham framework.

There exist several papers in the  literature making
unjustifiably strong conclusions  based upon analyses of conserved quantities,
such as  total momentum.  One example is that of Skobel'tsyn \cite{skobeltsyn73}
(that paper was discussed also in the review \cite{brevik79}, page
192). There is generally no definite answer about ``correctness" to be
obtained from this kind of analysis; one ends up performing consistency checks
of the formalism rather than giving real derivations.  In our opinion it is instead  from
experimental information that one can get reliable information
about appropriateness and usefulness (rather than ``correctness''), of the
various alternatives for electromagnetic momentum. 

\subsection{Second case: $b>x_0$}\label{sec:Ib}

Let us briefly return to Example I, but now allow the charge to lie within the cylindrical region in which the $z$-directed magnetic field is initially nonzero. Since the calculations are closely analogous to those in section \ref{sec:Ia}, we do not here give the full details of calculations. The contribution from the region $r\leq x_0$ is 
\be\label{GyIb}
  G_y^{M-} = \frac1{2}QB_0 x_0
\ee
as seen immediately from Eq.~\eqref{18}. To calculate the momentum contained within the annular cylindrical region $x_0<r<b$, we require the electric fields in this region, easily obtained by differentiation of \eqref{7}. Performing the integrals over $z$ and $\theta$, and summing over $m$ (only $m=\pm1$ contribute as before) we obtain the annulus momentum
\begin{align}
  G_y^{M+} =& \frac{4QB_0}{\pi L}\sum_{p=1}^\infty\sin^3\left(\frac{p\pi}{2}\right) I_1(pqx_0)\int_{x_0}^brdr\left\{p\left[K_1'(pqr)-\frac{K_1(pqR)}{I_1(pqR)}I_1'(pqr) \right]\right.\notag \\
  &+\left.\frac{1}{qr}\left[K_1(pqr)-\frac{K_1(pqR)}{I_1(pqR)}I_1(pqr) \right]\right\}.
\end{align}
Solving the integral over $r$ as above and taking the limit $L,R\gg b,x_0$ we obtain
\be
  G_y^{M+} = \lim_{R,L\to \infty}\frac{QB_0x_0(b^2-x_0^2)}{L^2}\Psi(R/L) \buildrel{L\to \infty}\over{\to} 0.
\ee
[The function $\Psi(x)= \sum_{l=1}^\infty (-1)^l(2l-1)^3K_1[(2l-1)x]/I_1[(2l-1)x]$ exists for $x>0$].

We thus see that there is no contribution of momentum from the annulus of radii $>x_0$ and the total momentum is given by \eqref{GyIb}. Note that this also holds in the limit $b\to \infty$, a point we will discuss later. One sees almost immediately that the total momentum is again transferred to the line charge and is conserved.

\section{Example II: Infinite dielectric cylinder, with external parallel line charge}

\begin{figure}[ht]
  \begin{center}
    \includegraphics[width=3in]{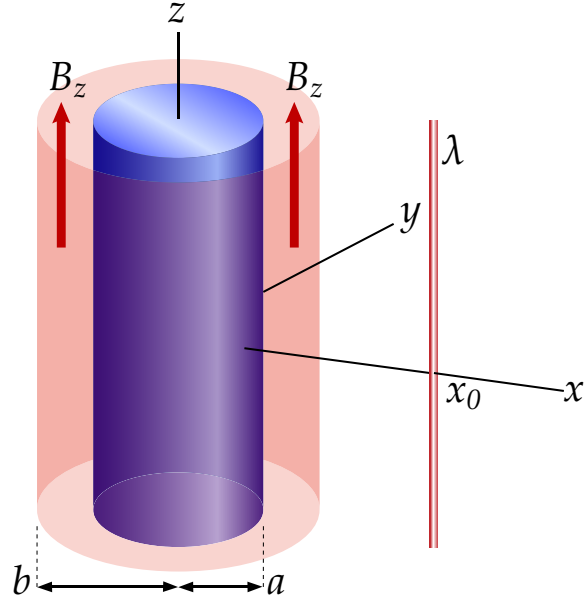}
    \caption{A line charge $\lambda$ per unit length outside an infinitely long dielectric cylinder of radius $a$, positioned coaxially with a magnetic field confined at radii $\leq b$. The case $a<b<x_0$ (second case) is shown here.}
    \label{fig3}
  \end{center}
\end{figure}

Consider as next example a dielectric nonmagnetic cylinder of radius $a$, of infinite length and permittivity $\varepsilon$, as shown in Fig.~\ref{fig3}. We employ cylindrical coordinates $r,\theta, z$, where the $z$ axis coincides with the cylinder's symmetry axis. In the outside region $r>a$ we assume that there is an infinite parallel line, of charge $\lambda$ per unit length, parallel to the cylinder. The line passes through the $xy$ plane at position $x=x_0, \,\theta=0$. The electric field lies entirely in the $xy$ plane, ${\bf E}=E_r\,\hat{\bf r}+E_\theta \,\hattheta$. For times $t<0$ there is  a uniform magnetic field $B_0\,\bf{\hat z}$ imposed in the $z$ direction. We will assume as in Example I that the magnetic field is restricted to radii smaller than a radius $b$. The three cases $b\leq a$ and $a<b<x_0$ and $b>x_0$ will be treated separately.
As before, we assume that the magnetic field decays for $t>0$.

We shall examine the momentum balance in the transverse $y$ direction, per unit length of the cylinder. As the length is infinite, this means that we are effectively dealing with a closed physical system.

\subsection{First case: $b\leq a$}

Consider  the internal electric field ${\bf E}^-$. The potential $\Phi^-$ is the same as that produced in a homogeneous medium (with permittivity $\varepsilon$) by a fictitious charge $2\lambda \varepsilon/(\varepsilon+1)$ placed at $x_0$ \cite{landau84}:
\begin{equation}
\Phi^-=-\frac{\lambda}{2\pi \varepsilon_0}\,\frac{1}{\varepsilon +1}\,\ln \left[r^2-2rx_0 \cos\theta +x_0^2\right]. \label{4.1}
\end{equation}
It leads to
\begin{subequations}\label{IIfields}
\begin{align}
  E_r^-=&\frac{\lambda}{\pi \varepsilon_0}\frac{1}{\varepsilon+1}\frac{r-x_0\cos \theta}{r^2-2rx_0\cos \theta +x_0^2}, \label{4.2} \\
  E_\theta^-=&\frac{\lambda}{\pi \varepsilon_0}\,\frac{1}{\varepsilon+1}\,\frac{x_0\sin \theta}{r^2-2rx_0\cos \theta +x_0^2}. \label{4.3}
\end{align}
\end{subequations}
The $y$ component of the interior momentum density is
\begin{equation}
  g_y^{M-}=-B_0(D_r^-\cos \theta-D_\theta^-\sin\theta), \label{4.4}
\end{equation}
from which it follows that
\begin{equation}
  G_y^{M-}=\int_0^{2\pi}d\theta \int_0^b  rdrg_y^{M-}. \label{4.5}
\end{equation}
We get
\begin{equation}
  G_y^{M-}=\frac{\lambda B_0}{\pi x_0}\frac{\varepsilon}{\varepsilon+1}\int_0^{2\pi}d\theta \int_0^b rdr\frac{1-(r/x_0)\cos \theta}{1-2(r/x_0)\cos \theta +(r/x_0)^2}. \label{4.7}
\end{equation}
Recognising the Fourier series
\begin{equation}
  \frac{1-s\cos \theta}{1-2s\cos \theta+s^2 }=\sum_{n=0}^\infty s^n \cos n\theta, \quad  s<1, \label{4.8}
\end{equation}
we see that the integral over $\theta$ in Eq.~(\ref{4.7}) gets contribution only from $n=0$. Thus
\begin{equation}
  G_y^{M-}=\frac{\lambda B_0b^2}{x_0} \frac{\varepsilon}{\varepsilon+1}. \label{4.9}
\end{equation}
One may note that in the limit $\varepsilon \to 1$ this expression becomes identical to (\ref{4}) and (\ref{18}), with $\lambda$ replacing the point charge $Q$.

Consider now the magnetic field decay period $t>0$, in which we have to include the induced electric field
$E_\theta ^{\rm induced}=-b^2{\dot B}/2r$ for $r\geq b$.  There is no electromagnetic force acting in the homogeneous interior; all forces are surface forces. Such forces can  be calculated by taking the difference between the radial components of the Maxwell stress tensor on the outside and the inside. However, it is in our case more convenient to start from the fact that the  force density in the surface layer is $-(\varepsilon_0/2)E^2{\bf \nabla}\varepsilon$.  By integrating this across the surface layer and using electromagnetic boundary conditions we can express the surface force density $\bdsigma$ as
\begin{equation}
  \bdsigma(r,\theta)=\frac{1}{2}\varepsilon_0(\varepsilon-1)\left[E_\theta^2(r,\theta)+\varepsilon E_r^2(r,\theta)\right]\Big|_{a-} \hat{ \bf r}, \label{26}
\end{equation}
thus showing that we need the interior fields only. The transverse $y$ component is
\begin{equation}
  \sigma_y(\theta)=\frac{1}{2}\varepsilon_0(\varepsilon-1)[E_\theta^2(a^-,\theta)+\varepsilon E_r^2(a^-,\theta)] \sin\theta, \label{4.20}
\end{equation}
where the internal fields 
are given by \eqref{IIfields} with the addition of $E_\theta ^{\rm induced}$ in the $\theta$ component.
When inserting these fields into the expression for the $y$ component of force on the cylinder,
\begin{equation}
  F_y^{\rm cylinder}=a\int_0^{2\pi}d\theta\sigma_y(a,\theta), \label{4.23}
\end{equation}
we do not need to evaluate all the individual terms since only those containing $\dot B$ can contribute. Indeed the only contribution comes from the cross term in  $(E_\theta^-)^2$, and we obtain
\begin{align}
  F_y^{\rm cylinder}=&-\frac{\lambda\dot{B}b^2}{2\pi x_0}\,\frac{\varepsilon-1}{\varepsilon+1}\int_0^{2\pi}\frac{d\theta\sin^2\theta }{1-2(a/x_0)\cos\theta+(a/x_0)^2}\notag \\
  =&-\frac{\lambda\dot{B}b^2}{2x_0}\frac{\varepsilon-1}{\varepsilon+1}, \label{4.24}
\end{align}
where we use the Fourier series
\be
  \frac{\sin\theta}{1-2s\cos\theta + s^2}=\sum_{n=1}^\infty s^{n-1} \sin n\theta
\ee
for $s<1$, which upon insertion into our integral contributes only for $n=1$.
This leads to the momentum transfer
\begin{equation}
  P_y^{\rm cylinder}=\int_0^\infty dt F_y^{\rm cylinder}=\frac{\lambda B_0b^2}{2x_0}\frac{\varepsilon-1}{\varepsilon+1}. \label{4.25}
\end{equation}

In addition to the force on the cylinder surface there is also an azimuthal force $F_y^{\rm line}=\lambda E_\theta^{\rm induced}$ acting on the line charge $\lambda$ during the decay period. The induced field is again $E_\theta^{\rm induced}=-\frac{b^2}{2r}\dot B$ so, with $r=x_0$,
\begin{equation}
  F_y^{\rm line}=-\frac{\lambda \dot{B}b^2}{2x_0}, \label{4.27}
\end{equation}
leading to the impulse transfer
\begin{equation}
  P_y^{\rm line}=\int_0^\infty dt F_y^{\rm line}=\frac{\lambda B_0 b^2}{2x_0}. \label{4.28}
\end{equation}
The total impulse transfer to the system during the decay period is accordingly
\begin{equation}
  P_y=P_y^{\rm cylinder}+P_y^{\rm line}=\frac{\lambda B_0 b^2}{x_0}\,\frac{\varepsilon}{\varepsilon+1}. \label{4.29}
\end{equation}
This expression is seen to be the same as Eq.~(\ref{4.9}). The momentum balance is thus verified, within the framework of the Minkowski formulation. To check the momentum balance in the Abraham case, one would have to take into account the Abraham force acting on the interior of the medium during the decay period, similarly as we did above.

\subsection{Second case: $a<b<x_0$}

For the case $b>a$ we must also consider the electric field from the line source in the region $r>a$.
The potential $\Phi^+$ is that produced in a vacuum by the actual charge $\lambda$, plus that of two fictitious line charges, one of magnitude $\lambda'=-\lambda (\varepsilon-1)/(\varepsilon+1)$ located at the position $x_1=a^2/x_0$, the other of magnitude $-\lambda'$ located at the origin \cite{landau84}, wherewith:
\begin{align}
  \Phi^+=&-\frac{\lambda}{4\pi \varepsilon_0} \left[\ln (r^2-2rx_0 \cos\theta +x_0^2)-\frac{\varepsilon-1}{\varepsilon+1}\ln (r^2-2rx_1 \cos\theta +x_1^{2}) \right.\notag \\
  &+\left.\frac{\varepsilon-1}{\varepsilon+1}\ln r^2 \right].
\end{align}
This gives the fields
\begin{align*}
  E_r^+ =& \frac{\lambda}{2\pi\varepsilon_0}\left[\frac{r-x_0\cos\theta}{r^2-2rx_0\cos\theta+x_0^2}-\frac{\varepsilon-1}{\varepsilon+1}\left(\frac{r-x_1\cos\theta}{r^2-2rx_1\cos\theta+x_1^2}-\frac{1}{r}\right)\right];\\
  E_\theta^+ =& \frac{\lambda}{2\pi\varepsilon_0}\left[\frac{x_0\sin\theta}{r^2-2rx_0\cos\theta+x_0^2}-\frac{\varepsilon-1}{\varepsilon+1}\frac{x_1\sin\theta}{r^2-2rx_1\cos\theta+x_1^2}\right].
\end{align*}
The resulting momentum density is found from \eqref{4.4} to be
\begin{align}
  g_y^{M+}(r,\theta) =& -\frac{\lambda B_0}{2\pi}\left[\frac{r\cos\theta-x_0}{r^2-2rx_0\cos\theta+x_0^2}-\frac{\varepsilon-1}{\varepsilon+1}\frac{r\cos\theta-x_1 }{r^2-2rx_1\cos\theta+x_1^2} \right.\notag \\
  &+\left.\frac{\varepsilon-1}{\varepsilon+1}\frac{\cos\theta}{r}\right]\label{gyAnnular}
\end{align}
where $\beta_1=b/x_1$ and $\tilde{r}_1=r/b$. The contribution to $G_y^{M+}$ is now
\be
  G_y^{M+} = \int_a^b rdr \int_0^{2\pi}d\theta g_y^{M+}(r,\theta).
\ee
The first term of (\ref{gyAnnular}) may be integrated over $\theta$ as before, using (\ref{4.8}). Since $x_1<r$ in the annular region between radii $a$ and $b$, however, we must use a different series expansion for the second term,
\begin{align}
  \frac{\cos\theta -s}{1-2s\cos \theta+s^2} =& \sum_{n=1}^\infty s^{2n-1}\cos 2n\theta + \cos\theta\sum_{n=0}^\infty(-1)^ns^{2n}\left[1-2\sum_{k=1}^n\cos 2k\theta\right]\notag\\
  =& \cos \theta + s \cos 2\theta - s^2\cos\theta(1-2\cos 2\theta)+ s^3\cos 4\theta+...\label{Taylor2}
\end{align}
for $s<1$, from which we see that the second term of \eqref{gyAnnular} gives no contribution when integrated from $\theta=0$ to $2\pi$. The same is clearly the case for the last term in Eq.~\eqref{gyAnnular}, so that only the first term of \eqref{gyAnnular} contributes;
\begin{align}
  G_y^{M+}=& \int_a^b rdr\int_0^{2\pi}d\theta g_y^{M+}(r, \theta)=\frac{\lambda B_0(b^2-a^2)}{2x_0}.
\end{align}
After inserting the momentum from the internal region from (\ref{4.9}) with $b\to a$, we find the total field momentum to be in the Minkowski formalism
\be
  G_y = G_y^{M-}+ G_y^{M+}=\frac{\lambda B_0}{2x_0}\left(b^2+\frac{\varepsilon-1}{\varepsilon+1}a^2\right).\label{Gybga}
\ee

We again turn to the period in which the magnetic field decays. The momentum transfer to the line charge and cylinder, respectively, are simply evaluated just like before, to yield
\begin{align}
  P_y^{\rm line}=&\frac{\lambda B_0 b^2}{2x_0};\\
  P_y^{\rm cylinder}=&\frac{\lambda B_0a^2}{2x_0}\frac{\varepsilon-1}{\varepsilon+1}.
\end{align}
Adding $P_y^{\rm line}$ and $P_y^{\rm cylinder}$ we retrieve the expression \eqref{Gybga}, and the momentum conservation is once again ascertained.

\subsection{Third case: $b>x_0$}

Consider finally the case where the line source lies within the volume in which the magnetic field is present. The momentum density inside the sphere is again given by \eqref{4.9} with $b\to a$, and the momentum density for $r>a$ again given by \eqref{gyAnnular}. As in the previous case, only the first term of \eqref{gyAnnular} contributes. Similar to the calculation in the previous section, different Taylor expansions of the fraction $(r\cos\theta -x_0)(r^2-2r x_0\cos \theta+x_0^2)$ must be employed depending on whether $r<x_0$ or the other way around: the series' \eqref{4.8} and \eqref{Taylor2}, respectively. The latter of these, as previously argued, gives no contribution after integration with respect to $\theta$, so we conclude that the annular region $x_0<r<b$, just as in the second case of Example I, carries no momentum, and the total field momentum for $t<0$ is found as
\be\label{GyIIc}
  G_y =\frac{\lambda B_0}{2x_0}\left(x_0^2+\frac{\varepsilon-1}{\varepsilon+1}a^2\right).
\ee
One easily verifies that the to terms in the paranthesis exactly equal the momentum transferred to the line source and dielectric cylinder, respectively. Note once again that this result holds also in the limit where $b\to \infty$. Subtleties connected to this limit are discussed in section \ref{sec:conclusions}.

\section{Example III: Dielectric sphere in an infinite magnetic field}

In the above two examples, we had the opportunity to exploit the cylinder symmetry property.  As a final example we shall consider briefly also the analogous problem with a dielectric sphere of radius $a$, acted upon by an external charge $Q$, and also exposed to a uniform magnetic field which for reasons of tractability we shall assume to be of {\it infinite} extent. 

Employing usual spherical coordinates $r,\theta, \phi$ it is mathematically convenient to place the external charge $Q$ on the $z$ axis, at position $z_0>a$ and let the transverse uniform magnetic field be directed normal to the sphere-charge line, along the $x$ axis: $B_0 \,\hat{ \bf x}$ for times $t<0$. As mentioned, the  magnetic field is  present everywhere, on the inside as well as on the outside.

\begin{figure}[ht]
  \begin{center}
    \includegraphics[width=2.5in]{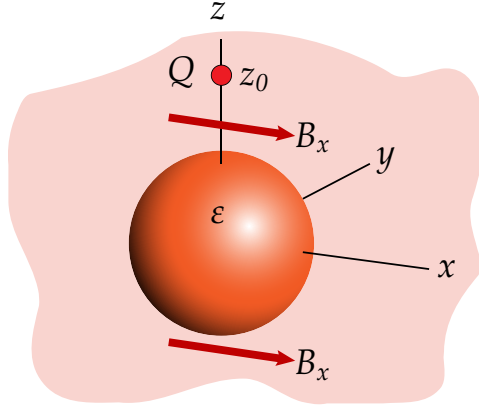}
    \caption{A point charge outside a dielectric sphere in an infinite $x$-directed magnetic field.}
    \label{fig4}
  \end{center}
\end{figure}

From Ref.~\cite{stratton41} we have for the potential on the inside
\begin{equation}
  \Phi^-=\frac{Q}{4\pi \varepsilon_0}\sum_{l=0}^\infty \frac{2l+1}{l\varepsilon+l+1}\,\frac{r^l}{z_0^{l+1}}\,P_l(\cos \theta), \label{27}
\end{equation}
giving rise to the field components
\begin{align}
  E_r^-=&-\frac{Q}{4\pi \varepsilon_0}\sum_{l=1}^\infty\frac{2l+1}{l\varepsilon+l+1}\, \frac{lr^{l-1}}{z_0^{l+1}}\,P_l(\cos\theta), \label{28}\\
  E_\theta^-=&\frac{Q}{4\pi\varepsilon_0}\sum_{l=1}^\infty \frac{2l+1}{l\varepsilon +l+1}\,\frac{r^{l-1}}{z_0^{l+1}}\,P_l'(\cos \theta)\sin\theta , \label{29}\\
  E_\phi^-=&0. \label{30}
\end{align}

The $y$ component of total momentum in the sphere is
\begin{equation}
  G_y^{M-}=2\pi \int_0^ar^2dr\int_0^\pi \sin \theta d\theta g_y^{M-}  , \label{31}
\end{equation}
where the density for $t<0$ is
\begin{equation}
  g_y^{M-}=B_0(D_r^- \cos \theta-D_\theta^- \sin \theta). \label{32}
\end{equation}
Inserting the expressions we get
\begin{align}
  G_y^{M-}=&-\frac{1}{2}QB_0\varepsilon a\sum_{l=1}^\infty \left( \frac{a}{z_0}\right)^{l+1}\frac{2l+1}{(l+2)(l\varepsilon+l+1)} \notag \\
  & \times \int_0^\pi [l\cos \theta P_l(\cos \theta)-\sin\theta P_l^1(\cos \theta)]\sin\theta  d\theta. \label{33}
\end{align}
Some algebra leads to the expression
\begin{equation}
  G_y^{M-}=-QB_0a\left( \frac{a}{z_0}\right)^2 \frac{\varepsilon}{\varepsilon+2}. \label{36}
\end{equation}

We have to consider the $y$ component of momentum from the fields on the outside as well. The potential  $\Phi^+$ in that region is  \cite{stratton41}
\begin{equation}
  \Phi^+=\frac{Q}{4\pi \varepsilon_0}\frac{1}{\varrho}-\frac{Q}{4\pi \varepsilon_0}(\varepsilon-1)\sum_{l=1}^\infty \frac{l}{l\varepsilon+l+1}\frac{a^{2l+1}}{z_0^{l+1}}\frac{P_l(\cos\theta)}{r^{l+1}}, \label{37A}
\end{equation}
which gives for the fields
\begin{align}
  E_r^+=&\frac{Q}{4\pi \varepsilon_0}\,\frac{r-z_0\cos\theta}{\varrho^3}-\frac{Q}{4\pi\varepsilon_0}(\varepsilon-1)\sum_{l=1}^\infty \frac{l(l+1)}{l\varepsilon+l+1}\frac{a^{2l+1}}{z_0^{l+1}}\frac{P_l(\cos\theta)}{r^{l+2}}, \label{38}\\
  E_\theta^+=&\frac{Q}{4\pi \varepsilon_0}\frac{z_0}{\varrho^3}\sin\theta+\frac{Q}{4\pi \varepsilon_0}(\varepsilon-1)\sum_{l=1}^\infty \frac{l}{l\varepsilon+l+1}\frac{a^{2l+1}}{z_0^{l+1}}\frac{P_l^1(\cos \theta)}{r^{l+2}}, \label{39}\\
  E_\phi^+=&0, \label{40}
\end{align}
where $\varrho=(r^2-2rz_0\cos\theta +z_0^2)^{1/2}$ is the
distance between $Q$ and the field point. These expressions are
inserted into the analog of Eq.~(\ref{32}) in the outer space (superscript minus replaced by plus). An
explicit calculation shows that the contribution from the medium-generated fields, i.e. the last
terms in Eqs.~(\ref{38}) and (\ref{39}), is actually zero. The contribution from the first terms, which stem from the point charge alone, are calculated by means of the relation 
\be
  \int_0^\pi d\theta \frac{\sin\theta(r\cos\theta-z_0)}{(r^2+z_0^2-2rz_0\cos\theta)^{3/2}} = -\frac{2}{z_0^2}\Theta(z_0-r)
\ee
where $\Theta(x)$ is the unit step function. We obtain the result
\be
  G_y^{M+} = -\frac{QB_0}{3z_0^2}(z_0^3-a^3)
\ee
so the total momentum of the electromagnetic fields before the decay period is
\be\label{GySphere}
  G_y = G_y^{M-}+G_y^{M+}= -\frac{2QB_0 a^3}{3z_0^2}\frac{\varepsilon-1}{\varepsilon+2} - \frac{QB_0 z_0}{3}.
\ee

A note is warranted at this point. The second term of \eqref{GySphere} is in a sense arbitrary: it refers to the absolute position of the point charge only, and not the sphere. The first term, in comparison, depends only on parameters which are independent of the coordinate system: the sphere radius and the ratio of the radius to the distance of the point charge from the sphere center. Would not then the second term change its value upon a linear transformation of coordinates, but not the first? And secondly: how can a term not referring to the sphere possibly be non-zero, when the remaining system of point charge and infinite field has perfect spherical symmetry?

In fact we have seen this paradox also in the previous chapters, although it was not discussed at the time. Both the field momentum \eqref{GyIb} for the particle in a cylindrical field and \eqref{GyIIc} are valid in the limit where the magnetic field fills all of space, although were one to take that limit before doing the calculations, one would get the result zero because of symmetry. Also here we must interpret the infinite magnetic field to be the limit where some confinement volume is taken to infinity. Depending on the original symmetry, the charge alone \emph{can} constitute a symmetry break (provided the limit of infinite field is taken at the end) and give rise to a field momentum different from zero, although the physical meaning of this momentum is not so obvious. We return to this point in Section \ref{sec:conclusions}.

Consider now the magnetic decay period, $t>0$. The impulse
transferred to the sphere, calculable from the fields at $r=a-$,
necessarily has to correspond to the surface force density
component
\begin{equation}
\sigma_y=\frac{1}{2}\varepsilon_0(\varepsilon-1)[E_\parallel^2+\varepsilon
E_r^2]\Big|_{a-}\sin\theta\sin\phi. \label{A.1}
\end{equation}
As discussed we assume azimuthal symmetry for
the induced electric field $ {\bf E}^{\rm induced}$  around the direction of $\bf B$. One finds
\begin{equation}
{\bf E}^{\rm induced}=-\frac{1}{2}\tilde{r}\dot{B} \,\bf \hat{x}\times \hat{\tilde{r}}, \label{A.2}
\end{equation}
where $\hat{\tilde{\mathbf{r}}}$ is the unit vector directed normal to, and away from, the $x$ axis [here $\tilde{r}=z/\sin\alpha$, and $\hat{\tilde{\mathbf{r}}}=\cos\alpha \hat{\mathbf{y}}+\sin\alpha \hat{\mathbf{z}}$ wherein  $\cot\alpha=y/z=\tan\theta\sin\phi$], implying the components
\begin{align}
  E_r^{\rm induced}=&0, \label{A.3}\\
  E_\theta^{\rm induced}=&\frac{1}{2}r\dot{B}\sin\phi, \label{A.4}\\
  E_\phi^{\rm induced}=&\frac{1}{2}r\dot{B}\cos \theta \cos \phi. \label{A.5}
\end{align}
Letting $r=a$ and adding these expressions to Eqs.~(\ref{28})-(\ref{30}) for the interior fields at $r=a-$, we obtain  after some algebra the following impulse transfer to the sphere
\begin{equation}
  P^{\rm sphere}_y=-\frac{1}{2}QB_0a\left(\frac{a}{z_0}\right)^2\frac{\varepsilon-1}{\varepsilon+2}. \label{A.7}
\end{equation}
The momentum transferred to the point charge is also easily calculated, assuming the charge stays close to $\theta=0$ throughout the decay period:
\be\label{PchargeSphere}
  P^{\rm charge}_y=-\frac1{2}QB_0z_0.
\ee
This is the same result obtained in Example I, case 2 above, Eq.~\eqref{18}, and for obvious reasons: in both cases the induced electric field is calculated from the single property of cylindrical symmetry of $\mathbf{B}$ around the $x$ axis. 

There appears to be a discrepancy present: unlike in the previous examples, the momentum transferred to charge and body does not equal that initially present in the electromagnetic fields. There is a ``missing'' momentum
\be\label{missingMomentum}
  \Delta G_y = -\frac1{6}QB_0a\left(\frac{a}{z_0}\right)^2\frac{\varepsilon-1}{\varepsilon+2}+\frac1{6}QB_0z_0.
\ee
Where has the momentum gone? The only possibility is that it has escaped the system towards infinity. We will now verify that this is indeed so.

This we do by calculating the net flux of $y$-directed momentum through a control surface which we place far from the sphere and point charge. We let the control surface be spherical for simplicity, with radius $R\gg a,z_0$. The flux of $y$-momentum through the surface is given as
\be
  \dot{G}_y = -\oint_{\mathrm{CS}} \hat{\mathbf{y}}\cdot \mathbb{T}\cdot d{\mathbf{A}}
  = -R^2\int_0^\pi d\theta \sin\theta\int_0^{2\pi}d\phi (\hat{\mathbf{y}}\cdot \mathbb{T}\cdot \hat{\mathbf{r}})(\theta,\phi)
\ee
where the stress tensor in vacuum is
\be
  \mathbb{T} = \varepsilon_0\mathbf{E}\otimes\mathbf{E}+\frac1{\mu_0}\mathbf{B}\otimes\mathbf{B}-\frac{\varepsilon_0}{2}(\mathbf{E}^2+c^2 \mathbf{B}^2)\mathbb{I}
\ee
where $\mathbb{I}$ is the unit matrix (Kronecker's delta). In vacuum the stress tensor is related to the (any) energy momentum $4-$tensor $S_{\mu\nu}$ by $T_{ik}=-S_{ik}$, latin letters denoting spatial dimensions. We obtain
\begin{align}
  \hat{\mathbf{y}}\cdot \mathbb{T}\cdot \hat{\mathbf{r}}=&
  \frac{\varepsilon_0}{2}[\sin\theta\sin\phi E_r^2 + 2\cos\theta\sin\phi E_\theta E_r - \sin\theta\sin\phi E_\theta^2 \notag\\
  &+ \cos\phi E_\phi E_r - \sin\theta\sin\phi E_\phi^2 - \sin\theta\sin\phi c^2B_0^2].
\end{align}
Only a few of these terms give a contribution after integration with respect to $\phi$. The field expressions are simplified after expanding in powers of $a/R$ and $z_0/R$
\begin{align}
  \int_0^{2\pi}d\phi(\hat{\mathbf{y}}\cdot \mathbb{T}\cdot\hat{\mathbf{r}})=&\frac{Q\dot{B}R}{8}\left[\frac{2\cos\theta}{R^2}+\frac{z_0}{2R^3}(3+5\cos 2\theta)\right.\notag \\
  &\left. -\frac{\varepsilon-1}{\varepsilon+2}\frac{a^3}{z_0^2 R^3}(4\cos^2\theta-\sin^2\theta)\right]+\mathcal{O}\left(\frac{a^4}{R^4},\frac{z_0^4}{R^4}\right).
\end{align}
The higher order terms will vanish when taking the radius $R$ to infinity and we need not consider them. Finally, performing the $\theta$ integral we obtain the momentum which has disappeared as flux during the decay period
\be
  G_y^\text{flux}=\int_0^\infty dt \dot{G}_y = \frac{Q B_0}{6}\left(z_0 - \frac{\varepsilon-1}{\varepsilon+2}\frac{a^3}{z_0^2}\right)
\ee
(recall: $\int_0^\infty dt \dot{B}=-B_0$) which implies that exactly all of the ``missing momentum'' $\Delta G_y$ in \eqref{missingMomentum} has disappeared to infinity. In other words, a momentum has been imparted on the infinite background medium during the decay period, not dissimilar to the effect of the Abraham force in section \ref{sec:abrahamMinkowski}. This striking result is discussed in the following.

\section{Discussion and conclusions}\label{sec:conclusions}

We have verified the momentum balance of electromagnetic momentum in decaying magnetic fields in the presence of dielectric media in three different configurations: first a point charge and a homogeneous magnetic field, the latter restricted to a cylindrical volume, embedded in an infinite dielectric medium, secondly a point charge in vacuum outside a dielectric cylinder aligned coaxially with the same magnetic field, and finally the case of a dielectric sphere and a point charge in the presence of an infinite and homogeneous magnetic field. While the momentum balance is simply and uniquely calculated in the first two examples, the third example is more subtle, and in that case we find that some of the momentum disappears to infinity as momentum flux, calculated by use of the energy-momentum tensor.

The fact that part of the initial momentum leaves the system is a striking find. The ``physicalness'' of the result is not so obvious, however, because the infinite extent of the magnetic field introduces arbitrariness, as we will now explain and discuss. First, however, note a second and similar paradox which appears upon comparing Examples I and III: In Eq.~\eqref{GySphere} we found one term which clearly refers to both sphere and charge, and another referring to the charge alone. It is natural, as discussed above, to interpret the second term as pertaining to the single charge and the (axisymmetric) magnetic field, and would if so be expected to have the same value even if the sphere were not present. Surprisingly, however, the latter term does \emph{not} equal that of just a charge inside a cylindrical magnetic field, Eq.~\eqref{GyIb}, even when the cylindrical field volume radius is taken to infinity. 

Both of these surprising observations are connected to the unphysical assumptions of a magnetic field which fills all of space, which we made in order to be able to perform explicit calculations. In fact, this limit is not unique -- it matters what the shape of the field volume is as the limit is taken. Note how, in this limit $b\to \infty$, the results Eq.~\eqref{GyIb} and the first term of (\ref{GyIIc}) (just like the second term of Eq.~\eqref{GySphere}) become arbitrary: once the magnetic field is \emph{infinite} in every direction, new symmetries emerge. We are now free to move and rotate the coordinate system freely. Then the formal momentum of the electromagnetic field of a point charge in an infinite magnetic field, which depends on the absolute coordinates of the charge, can take \emph{any real value}, depending on how the limit is taken!

\begin{figure}[ht]
  \begin{center}
    \includegraphics[width=5in]{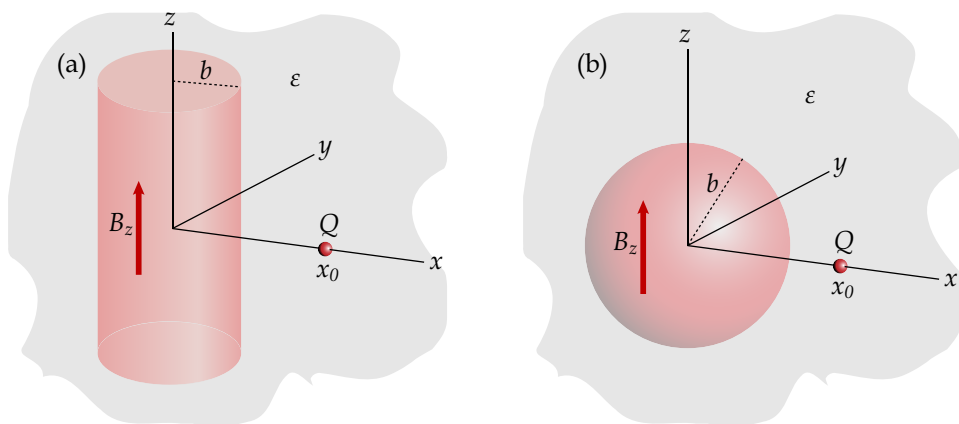}
    \caption{Gedanken experiment: two different ``field volumes'' in an infinite medium.}
    \label{fig5}
  \end{center}
\end{figure}

To illustrate even further, consider the following gedanken experiment, illustrated in Fig~\ref{fig5}. The first situation, Fig.~\ref{fig5}a, is the same as in Example I, the second, Fig.~\ref{fig5}b, is the hypothetical situation of a magnetic field $B_0 \hat{\mathbf{z}}$ confined to a spherical volume of radius $b$. The latter situation is clearly unphysical since magnetic field lines must be closed loops, yet let us ignore this fact for now. Using the potential of Eq.~\eqref{37A} we can calcluate (with a cyclic permutation of coordinates) the initial momentum contained in the spherical volume. We get
\begin{align*}
  G_y^M =& \frac{QB_0}{2}\times\left\{\begin{array}{cl}\frac{b^3}{z_0^2}, & b<z_0,\\ z_0, &b\geq z_0\end{array}\right.
  \text{ for the cylinder, and }\\
  G_y^M =& \frac{QB_0}{3}\times\left\{\begin{array}{cl}\frac{b^3}{z_0^2}, & b<z_0,\\ z_0, &b\geq z_0\end{array}\right.
  \text{ for the sphere.}
\end{align*}
The former is the result of Example I, the latter we recognize as the momentum term in the spherical example, Eq.~\eqref{GySphere}, which refers only to the charge. When $b\to \infty$ in both cases one obtains the magnetic field filling all of space, but formally the momentum arising from the point charge is different in the two cases. 

The discrepancy between the initial momentum and that transferred to the sphere and point charge in example III then has the following explanation: the former is calculated taking the infinite magnetic field to be the limit of the spherical field in Fig.~5, the latter taking the cylindrical limit. The difference, being purely a mathematical artifact, cannot give rise to physically measurable forces and is therefore absorbed by the infinite background medium. 

The subtlety relating to infinite magnetic field extent can be tackled by prescription: One may redefine the initial momentum of the initial fields by subtracting the ``unobservable'' amount which disappears as flux when the magnetic field decays (calculating both quantities in the same formalism). In Example III we then get the ``renormalized'' momentum $-\frac{QB_0 a^3}{2z_0^2}\frac{\varepsilon-1}{\varepsilon+2} - \frac{QB_0 z_0}{2}$, which has the ``cylinder'' prefactor $1/2$, which is satisfactory since the cylindrical limit in Fig.~\ref{fig5} is the more physical (such a field can be produced inside a large solenoid, the magnetic field lines closed at infinity). This ``renormalization'' procedure is closely analogous to the transfer between Minkowski versus Abraham momentum which we discussed in section \ref{sec:abrahamMinkowski}: there the Abraham momentum is obtained by subtracting from the Minkowski momentum the impulse transferred by the Abraham force during the decay period. It is tempting to state that the ``renormalized'' momentum is the more physical, although as the century-old Abraham-Minkowski controversy has shown that such statements are subtle. The momentum balance in the presence of an infinite magnetic field, and hence the formal consistency, is found to be in order with or without such ``renormalization'' as it should.

We emphasize once again that, while we have used the Minkowski formalism for electromagnetic momentum in matter in the examples calcuated in previous sections, the Abraham momentum could equally well have been made use of as discussed in section \ref{sec:abrahamMinkowski}. Considerations of conserved quantities cannot be used to support a particular energy-momentum tensor over another: such calculations are merely internal consistency checks which both the Abraham, Minkowski and other proposed energy-momentum tensors should all pass if calculations are carried out correctly. It is our belief that the appropriate criterion for distinguishing between tensors is whether they can explain experimental observations in a straightforward manner. A classic example of such an experiment is that of Walker and Lahoz \cite{walker75}. Another set-up was recently suggested by us \cite{brevik10}.

\end{document}